\documentclass[a4paper]{article}
\usepackage{ISCSLP2021}
\usepackage{multirow}
\usepackage{booktabs}
\usepackage{color}
\title{Context-aware RNNLM Rescoring for Conversational Speech Recognition}
\name{Kun Wei$^1$, Pengcheng Guo$^1$, Hang Lv$^1$,  Zhen Tu$^2$, Lei Xie$^1$*}

\address{
  $^1$Audio, Speech and Language Processing Group (ASLP@NPU), School of Computer Science, Northwestern Polytechnical University, Xian, China\\
  $^2$Zhuiyi Technology, Shenzhen, China}
\email{ethanwei@mail.nwpu.edu.cn, \{hanglv, pcguo, lxie\}@nwpu-aslp.org, jasontu@wezhuiyi.com}

\begin{document}

\maketitle
\begin{abstract}
Conversational speech recognition is regarded as a challenging task due to its free-style speaking and long-term contextual dependencies. Prior work has explored the modeling of long-range context through RNNLM rescoring with improved performance. To further take advantage of the persisted nature during a conversation, such as topics or speaker turn, we extend the rescoring procedure to a new context-aware manner. For RNNLM training, we capture the contextual dependencies by concatenating adjacent sentences with various tag words, such as speaker or intention information. For lattice rescoring, the lattice of adjacent sentences are also connected with the first-pass decoded result by tag words. Besides, we also adopt a selective concatenation strategy based on \textit{tf-idf}, making the best use of contextual similarity to improve transcription performance. Results on four different conversation test sets show that our approach yields up to 13.1\% and 6\% relative char-error-rate (CER) reduction compared with 1st-pass decoding and common lattice-rescoring, respectively.
\end{abstract}
\noindent\textbf{Index Terms}: conversational speech recognition, recurrent neural network language model, lattice-rescoring
\renewcommand{\thefootnote}{\fnsymbol{footnote}}
\footnotetext{* Lei Xie is the corresponding author.}
\section{Introduction}

With the development of deep neural networks, the performance of automatic speech recognition (ASR) on `ideal' scenarios has been advanced dramatically. However, ASR systems can be susceptible to performance degradation in real-life conversations, such as call-center or meeting transcription. Unlike news broadcast or audiobook data, conversations persist free-style spontaneous speech between several speakers (two and more). Although some common conversational speech recognition tasks, e.g., Switchboard~\cite{godfrey1992switchboard}, have achieved impressively low error rate \cite{han2017capio,kurata2017language}, transcribing real-world conversations are still challenging due to blended factors like heavy accents, noise interferences, speech disfluency as well as difficulty in the manual annotation. Since conversations usually have a long-term contextual dependency on nature, a straightforward idea is to capture these information in language modeling and decoding.

    A contextual ASR system can be obtained by modeling the long-term context in recurrent neural network based language models (RNNLMs) or adding additional context information to the neural network layers. To make use of contextual information, cache model was proposed to personalize the language models (LMs) based on recent language produced by the decoded speaker~\cite{kuhn1990cache, besling1995language, clarkson1997language, aleksic2015bringing, grave2016improving}. Dynamic evaluation fits models to the recent sequence history, allowing them to assign higher probabilities to re-occurring sequential patterns~\cite{krause2019dynamic}.  Another focus of modeling contextual information is LM rescoring~\cite{li2018recurrent, ogawa2019improved, liu2020contextualizing}. In \cite{velikovich2018semantic}, Velikovich \emph{et al}. proposed to identify and rescore semantically relevant entities or phrases in the word lattice.

    In recent years, RNNLMs have been proved to achieve SOTA performance and give significant improvement over the conventional back-off \emph{n}-gram in ASR and related tasks \cite{chen2017investigating,he2016training,sundermeyer2012lstm,mikolov2011rnnlm,lam2019gaussian}. The common idea to take advantage of RNNLMs for ASR tasks is a 2-pass decoding procedure. Firstly, a set of possible hypotheses are obtained by decoding on a precompiled \emph{n}-gram based graph. Then, we rescore these hypotheses through a more sophisticated neural-based language model. N-best list rescoring and lattice-rescoring are the most popular approaches for RNNLM-based rescoring. In this study, we mainly focus on the lattice-rescoring approach with the aim to boost performance on real-world conversational speech transcription.

    There has been plenty of work exploring the usage of RNNLMs lattice-rescoring. In~\cite{liu2016two}, Liu \textit{et al.} proposed two methods to speedup lattice-rescoring and compress the size of generated lattice. In~\cite{sundermeyer2014lattice}, Sundermeyer \textit{et al.} combined the previous lattice decoding work with long short-term memory (LSTM) neural network language models and also investigated a refined pruning technique. For capturing both past and future word contexts in language modeling, Chen \textit{et al.} proposed a more efficient structure succeeding word RNNLMs (su-RNNLMs), which uses a feedforward unit to model a finite number of succeeding future words~\cite{chen2017future}. In~\cite{khayrallah2017neural}, a stack-based lattice search algorithm was applied to constraint the search space in machine translation. In~\cite{xu2018pruned}, Xu \textit{et al.} further pruned the lattice by using a heuristic search to pick better histories. In addition to previous research mainly aimed at compressing the size of lattice, training a scenario-specific language model also becomes another focus of the community. In \cite{xiong2018microsoft,irie2019training}, the method of cross-sentences training is used to improve the language model performance.

    In this paper, we combine the ideas of scenario-specific language model and effective lattice rescoring in a more challenging task, conversational ASR, and propose a context-aware lattice rescoring method. Specifically, we first train an RNNLM by concatenating adjacent sentences with various tag words, such as separation identifier (SP), speaker label (SID) or intent label (INT). This explicit concatenation helps the language model to capture the contextual dependencies within a conversation, instead of a long sentence. When rescoring, we also connect the word lattice generated by the first decoding of the previous sentence with the lattice of current sentence. Tag words are used between every two lattices to mark the end position of the sentence and dock the speaker or intention information in RNNLM. Besides, a common similarity score based on \textit{tf-idf} (term frequency-inverse document frequency) can help in the selection of more suitable lattices to connect. Results on 4 different conversation test sets show that our approach yields up to 13.1\% and 6\% char-error-rate (CER) compared with 1st-pass decoding and lattice-rescoring relatively.

    The rest of this paper is organized as follows. In Section 2 and Section 3, we describe the idea of context-aware RNNLM and present the detail of the rescoring procedure of context-aware lattice rescoring. The experimental setup, results and analysis are described in Section 4. Finally, we conclude our work and discuss the possible future work in Section 5.
\section{Context-aware RNNLM}

\begin{figure}[htbp]
        \centering
        \includegraphics[width=\linewidth]{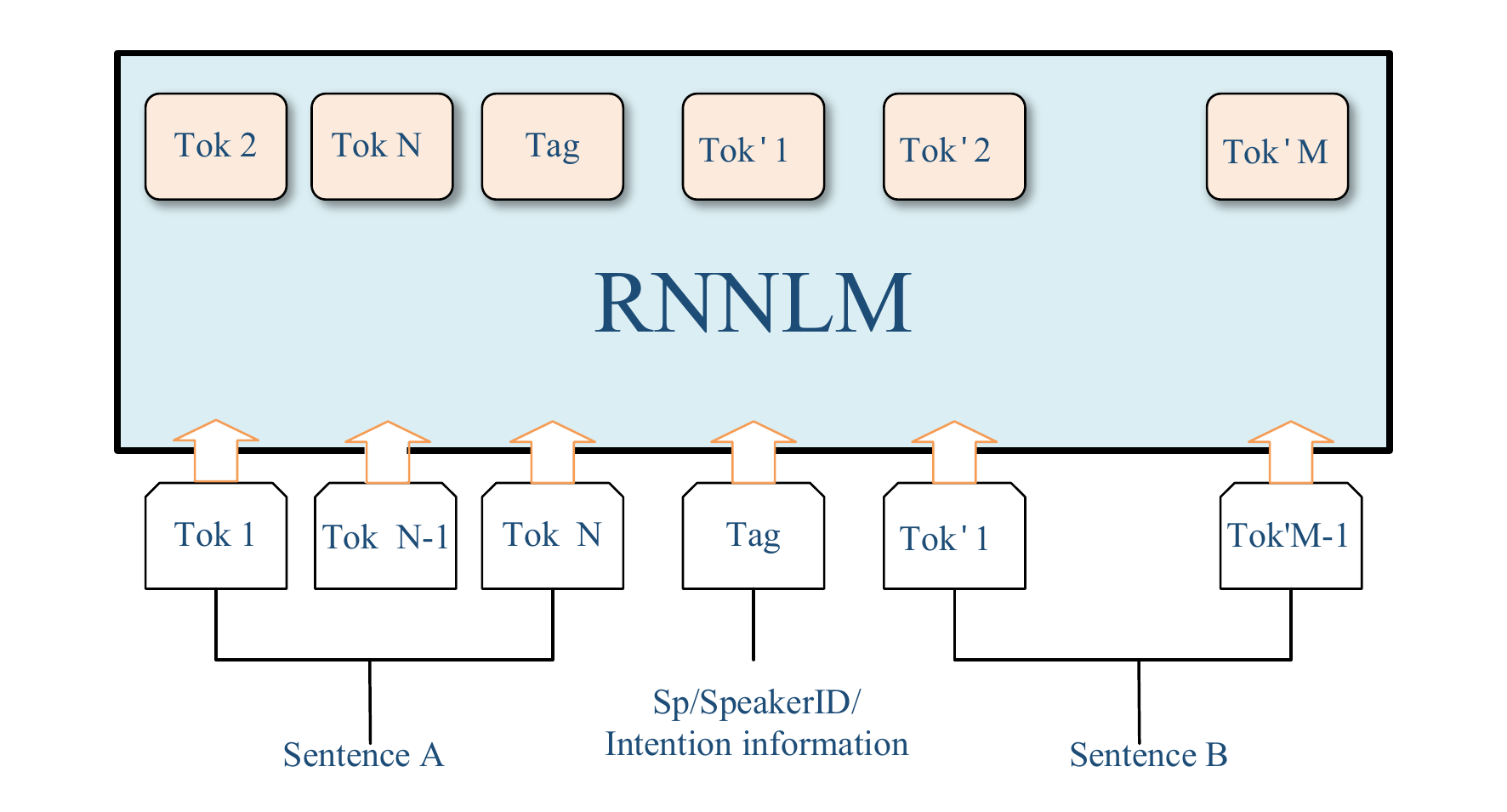}
        \caption{Schematic diagram of concatenating two sentences. The tag word can be a meaningless separation identifier (SP), speaker label (SID) or intention label (INT).}
        \label{fig:con_sen}
    \end{figure}

    In a conversational speech, multiple consecutive utterances are centered around the same common topics, which maintains a relationship between each other. In this case, it is natural to take account of the cross-sentence context during the modeling of language. There is no doubt that the cross-sentence language model can capture more contextual information, such as language styles and keywords, which may be ignored by conventional ones. Thus, we explore the concatenation of several consecutive sentences with tag words during the language model training, as shown in Fig. \ref{fig:con_sen}.

    We first cyclically concatenate the sentences in each dialogue and regard these long sequences as a new training set. The purpose of concatenating is to retain the sentence-level structure while making the back-propagation span of training beyond the sentence boundaries. It is conceivable that the concatenating length required for different topics needs to be optimized dynamically. Besides, the tag words between sentences should also carry more sentence related information instead of just as an ending identifier. Considering the difference in speaker and intent of each partial sentence, we alternatively replace the original separation identifier (SP) with speaker label (SID) or intent label (INT).

\section{Context-aware Lattice Rescoring}
We propose a context-aware lattice rescoring method based on a widely used pruned lattice rescoring approach. Before introducing our proposed method, we will firstly give a brief introduction about RNNLM lattice rescoring.

    \subsection{\emph{N}-gram approximation lattice-rescoring}
    Conceptually, RNNLM lattice-rescoring works by composing the lattice (named L) with a finite state acceptor (named F)~\cite{xu2018pruned}. F represents the difference between the \emph{n}-gram FST, which is already in the raw lattice, and the RNNLM. By composing L with F, we instantaneously remove part of the existing LM cost and add in new RNNLM costs. The F is generated on-demand by composing two deterministic finite state acceptors: one for the back-off \emph{n}-gram language model, and another is for the RNNLM.

    However, lattice-rescoring algorithm is not feasible in practice because the resulting lattice grows exponentially w.r.t the depth of the original lattice. Usually, an \emph{n}-gram approximation algorithm is used to limit the size of F. The algorithm works by merging history states in F that are the same in the (\emph{n} - 1) most recent words.

    \subsection{Pruned lattice-rescoring}

    The pruning-based lattice rescoring method utilizes a heuristic similar to that of A* search. This heuristic search greatly reduces the runtime of the algorithm and helps picking better histories when expanding the lattices, resulting in better ASR performances. The method is a kind of pruned FST composition algorithm. We assume that L is an epsilon-free acyclic lattice and F is an an epsilon-free (it means there can be at most one arc leaving from state $b$ in F that matchs arc leaving state $a$ in L) deterministic acceptor. When composing $C = L \circ F$, each state $c$ in the output FST corresponds to a pair of states $(a; b)$, and each arc in the output FST corresponds to a pair of arcs where the output label of the arc in L matches with the input label of the state in F. The total cost of state $c$ in the lattice L is computed using Viterbi “forward” and “backward” algorithms, which can be formulated as $\alpha (c) + \beta (c)$. Specifically, $\alpha(c)$ is the cost of the best path from the start-state to $c$, $H(c)$ is the final heuristic result, and $\beta(a)$ is the cost of the best path from $a$ to a final-state in L.
    \begin{equation}
      H(c) = \alpha(c) + \beta(a) +  \delta(c)
    \label{eq1}
    \end{equation}
     This method use $\beta(a) +  \delta(c)$ to approximate representation $\beta(c)$, the theoretically optimal path, because if we have not already expanded a state $c$, its $\beta(c)$ will always be $+\infty$. $\delta(c)$ is a correction factor that reflects how much we expect $\beta(c)$ to differ from $\beta(a)$. The reader can refer \cite{xu2018pruned} for a more detail explanation.

    \subsection{Context-aware Lattice-rescoring}

    The modeling capabilities of the pruned lattice-rescoring method is limited due to the \emph{n}-gram approximation algorithm. We introduce topic-based separator tag words between every two adjacent lattices to make better use of the context information on the same topic.  The topic-based tag is consistent with the language model, can be SP, SID or INT. Then the transition probability of this tag word from RNNLMs can help to capture more topic information from contextual sentences.

\begin{figure}[h]
        \centering
        \includegraphics[width=\linewidth]{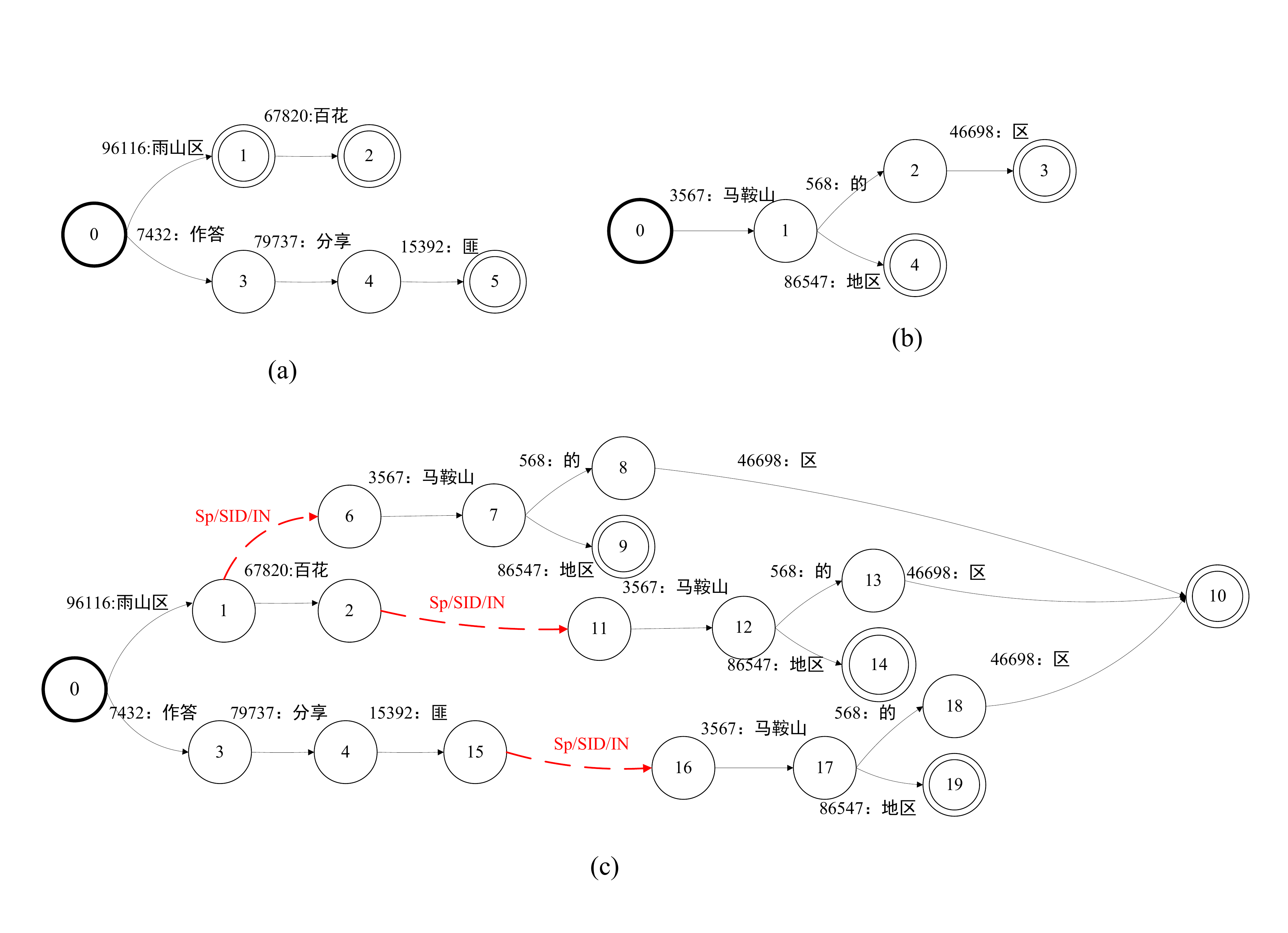}
        \caption{Details of lattice concatenation.  (a) and (b) are the compact lattices before concatenation and (c) is concatenated compact lattice. The black circle represents the start node, the ring represents the end node, and SP/SID/SIN means different tag words.}
        \label{fig:fst}
    \end{figure}

    For our proposed context-aware lattice rescoring method, the topic-based tag word is inserted between the previous one contextual lattice $L_{pre}$ and the current lattice $L_{cur}$ to form a new connected lattice $L_{con}$. When connecting lattices, we assign the final state's transition probability and the language model score of the previous lattice $L_{pre}$ to the arc which connect the original final state and the tag word of new connected lattice $L_{con}$. The same operation is used on the arc of the current lattice $L_{cur}$ from the starting node to those states which are reachable in one transfer, and the weights on other arcs remain unchanged, as described Eq~(\ref{eq2}) and Eq.~(\ref{eq3}):
    \begin{equation}
      W_{f\_s}(L_{con}) = W_{F}(L_{pre})
      \label{eq2}
    \end{equation}
    \begin{equation}
      W_{s\_t*}(L_{con}) = W_{S\_t*}(L_{cur})
      \label{eq3}
    \end{equation}
    where $W_{f\_s}$ means the weight from state $f$ to tag state $s$, which contains transition probability and the language model score, and $F$ means the final state of $L_{pre}$. $s$ denotes the tag state in $L_{con}$, $S$ indicates the start state, and $t*$ means those states which are reachable in one transfer in $L_{cur}$. To save the operation cost and ensure that the value of the probability transfer is not changed, we splice $n$-1 states of $L_{cur}$ after each end state of $L_{pre}$,  and these paths are transferred according to the different number of $n$-gram approximations starting from $n$ states. Figure \ref{fig:fst} is an example when $n$ is equal to 4.

\section{Experiments}

\subsection{Datasets and Setup}

    We conduct experiments on 2 different Mandarin conversational speech corpus: 1) Yispeech, a call center conversation dataset, which consists of more than 20 different financial domains; 2) HKUST~\cite{liu2006hkust}, a spontaneous telephone conversation dataset and each conversation are centered around a specific topic. For Yispeech, 3263 hours data are used to train the acoustic model, while 3 test sets with specific domains are used for evaluation, namely ZA, RQ, and DD. There exist domain overlaps in training and test set but the training set covers more domains. The LM is trained by 3.4G financial domain text collected from the Internet. For HKUST, the LM is trained by the transcriptions of training set. Detail descriptions of the dataset are shown in Table~\ref{tab:dataset}.

    \begin{table}[htbp]
    \caption{The statistics of of different conversational speech corpus.}
    \centering
    \begin{tabular}{l|l|l|l}
    \toprule[1pt]
    Corpus                    & Training set (hrs)    & \multicolumn{2}{l}{Testing set (hrs/convers)} \\ \hline
    \multirow{3}{*}{Yispeech} & \multirow{3}{*}{3263} & ZA Subset             & 1.89/2419             \\ \cline{3-4}
                              &                       & RQ Subset             & 2.06/4834             \\ \cline{3-4}
                              &                       & DD Subset             & 2.60/4721             \\ \hline
    HKUST                     & 200                   & \multicolumn{2}{l}{12.34/5413}               \\ \bottomrule[1pt]
    \end{tabular}
    \label{tab:dataset}
    \end{table}

     We use time-delay neural networks (TDNNs) trained with lattice-free MMI~\cite{povey2016purely} to build acoustic models in all experiments. The 43-dimensional log-fbank plus 100-dimensional i-vectors are extracted as the input features. The acoustic model consists of 12-layer TDNN. Each TDNN layer has 1280 hidden units followed by a batch normalization layer and a rectified linear units (ReLU) activation function. The output dimension of the acoustic model is 1563. For RNNLM rescoring, we use a 3-layer LSTM network to train the RNNLM, each layer of the RNNLM has 256 hidden units. The lattice is first generated by a 4-gram language model further rescored as described in~\cite{xu2018neural}. We adopt CER as the evaluation metric for different settings of RNNLMs and lattice rescoring. All experiments are based on Kaldi~\cite{povey2011kaldi}.

\subsection{Results}

\subsubsection{Context-aware RNNLM}

    We first investigate the effect of different concatenation numbers of the sentence in RNNLM training. Here, we use the common lattice rescoring method without lattice concatenation. When comparing no-concatenating, concatenating 2, 4, and all texts (a most extreme case) with SP, we find the setup of 4 achieves the best performance, so we use concatenating 4 sentence without concatenate tags as our baseline (named w/o Tags). All the following experiments will set the number to 4. Then,  we insert different tag word when perform sentence concatenation, such as SP, SID and INT. All experments use the lattice rescoring method without context informartion. It can be seen from Table \ref{tab:result1} that the method of SID achieves the best results on all test sets. On HKUST task, since the training set of RNNLM is small and covers a wide range of fields, both RNNLM without concatenation and concatenation without tags performs worse than 1st-pass decoding.
    \begin{table}[htbp]
    \centering
    \caption{The CER (\%) performance of 1st-pass decoding, rescoring use RNNLM without tags, and RNNLM with SP, SID,INT tags in traing set text on different conversational test set.}
    \begin{tabular}{l  l  l  l l l}
    \toprule[1pt]
        \textbf{Test set} & \textbf{1-pass} & \textbf{w/o Tags} & \textbf{SP} & \textbf{SID} & \textbf{INT} \\ \hline
        ZA & 20.01 & 20.23 & 18.63 & \textbf{18.51} & 19.47 \\
        RQ & 13.13 & 13.07 & 12.96 & \textbf{12.89} & -\\
        DD & 13.50 & 13.73 & 12.87 & \textbf{12.85} & - \\ \hline
        HKUST & 23.71 & 27.65 & 22.92 & \textbf{22.72} & - \\ \bottomrule[1pt]
    \end{tabular}

    \label{tab:result1}
    \end{table}
\renewcommand{\thefootnote}{\arabic{footnote}}
\subsubsection{Context-aware Lattice-rescoring}
    We find that the method of SID achieves the best results in context-aware RNNLM experiments in Table \ref{tab:result1}. So in the following experiments, we use the SID as the tag word in the training of context-aware RNNLM. Table~\ref{tab:result2} shows the CER performance of the conventional lattice rescoring method and our proposed context-aware lattice rescoring with different tag words. The \emph{w/o Tags} column use the SID RNNLM without lattice concatenation, so the results is the same as the SID column in Table \ref{tab:result1}. Here we only connect two adjacent lattice as Fig. \ref{fig:con_sen} because the computational complexity will increase exponentially with the increase of connected lattices.  When using SP as tag words, the CERs are 18.28\% and 12.78\% on ZA and RQ sets, which achieves 2.3\% and 0.9\% relative improvement over the conventional lattice rescoring. However, we don't get better results on the other two test sets. We further explore to replace SP with SID or INT, as presented in the 5th and 6th columns in the table. For SID, we add the ground truth speaker ID of the previous lattice as an extra token when concatenating. The SID use two different tag words to distinguish the two speakers in each topic. The results of SID are a little worse than both baseline and SP, which indicates that the speaker's information does not play an important role for rescoring. For INT, although speaker label is useful in RNNLM as described in Section 4.2.1, since we can't access the ground truth intent labels, we adopt the predictions of an intent classifier as the tag words for concatenating, which can be seen as `soft' intent labels. To obtain a robust intent classifier, we first divide all transcriptions of ZA set into 27 different categories and use them to fine-tune a pre-trained BERT-based LM model. \footnote{The pre-trained BERT-based LM model is an open-source model trained on a large text corpus like Wikipedia, which can be download in https://github.com/google-research/bert.} The intent classifier achieves 67\% prediction accuracy on the ZA set. When replacing the SP with our INT, we find the results are still worse than the baseline and SP-based method. It should also be noted that we only conduct INT experiments on one set, because only the ZA set has intent labels provided.


    \begin{table}[htbp]
    \centering
    \caption{The CER (\%) performance of 1st-pass decoding, base lattice-rescoring, and context-aware lattice-rescoring on different conversational test set.}
    \begin{tabular}{l  l  l  l l l}
    \toprule[1pt]
        \textbf{Test set} & \textbf{1-pass} & \textbf{w/o Tags} & \textbf{SP} & \textbf{SID} & \textbf{INT} \\ \hline
        ZA & 20.01 & 18.51 & \textbf{18.28} & 18.63 & 18.81 \\
        RQ & 13.13 & 12.89 & \textbf{12.78} & 12.89 & -\\
        DD & 13.50 & \textbf{12.85} & 13.08 & 13.21 & - \\ \hline
        HKUST & 23.71 & \textbf{22.72} & 23.09 & 25.23 & - \\ \bottomrule[1pt]
    \end{tabular}

    \label{tab:result2}
    \end{table}


\subsubsection{Selective lattice concatenation}
    Comparing the results of baseline and proposed methods, we find it will be more useful to connect 2 related sentences, e.g., the sentences have same keywords, than 2 absolutely different sentences. To make better use of the contextual information, we also propose a selective method for lattice concatenation. We use the \emph{tf-idf} (term frequency-inverse document frequency), a statistical method that uses word frequency and inverse text frequency to assess the importance of a word to a document, to measure the similarity of adjacent sentences. Firstly, we generate 1-best results from the 1-st pass decoding lattice. Then, we compute the \emph{tf-idf} of each adjacent sentence as their similarity score. The similarity score is ranged from 0 to 1 and the larger the score is, the more similar the two sentences are. During the lattice rescoring, we only concatenate the lattices whose similarity score are above a specific threshold. Table \ref{tab:tfidf} presents the results of lattice concatenate with different thresholds.

    The selective concatenation achieves the best results than concatenating all lattices and the baseline method on all 4 test sets. On ZA, DD, and HKUST sets, 0.0 seems to be a better choice of threshold since the score of around two-thirds of sentence pairs in these sets is 0.0, and a lower threshold may provide as much as possible contextual information. For the RQ set, a larger threshold brings more gain because the language style of each conversation is very different. Thus, a larger threshold could effectively prevent such unrelated contextual information, which may easily lead ambiguities in recognition.

    \begin{table}[htbp]
    \centering
    \footnotesize
    \caption{The CER (\%) performance of selective concatenation with different similarity score, \textbf{none} means don't use the similarity score to select lattice.}
    \begin{tabular}{lllllllll}
    \toprule[1pt]
    \multicolumn{1}{c}{} &
      \textbf{none} &
      \multicolumn{1}{c}{\textbf{$>$0.0}} &
      \multicolumn{1}{c}{\textbf{$>$0.1}} &
      \multicolumn{1}{c}{\textbf{$>$0.3}} &
      \textbf{$>$0.5} &
      \textbf{$>$0.9} \\ \hline
    \textbf{ZA} & 18.28 & \textbf{17.40} & 17.86 & 18.13  & 18.11 & 18.11 \\
    \textbf{RQ}  & 12.78 & 12.70 & 12.69  & 12.67 & 12.67 & \textbf{12.66} \\
    \textbf{DD}   & 13.08 & \textbf{12.71} & 12.89 & 12.88 & 12.83 & 12.83 \\ \hline
    \textbf{HKUST}  & 23.09 & \textbf{22.63} & 22.75  & 22.76  & 22.74 & 22.74 \\ \bottomrule[1pt]
    \end{tabular}
    \label{tab:tfidf}
    \end{table}

\subsubsection{Analysis}
        To understand which contextual information is more important for RNNLMs and lattice-rescoring, we also make an analysis based on the above results. For RNNLMs, we find that concatenate sentences with their speaker labels achieves the best performance. It's reasonable to explicitly add speaker information to conversational speech recognition, cause such information may help the RNNLMs to capture speaker turns and speaking styles. A large concatenate number of sentences may bring more contextual information, while also increase redundant dependencies. Here we find 4 is the optimal setup for RNNLMs to get a trade-off.


        We compare our results with the results of the basic rescoring algorithm (Table \ref{tab:result2}) and found that our method achieves the purpose of using contextual information (Table \ref{tab:egs}). We can see that the basic lattice rescoring method results are huixin and No.89 , the context-aware lattice rescoring method get the information of previous one sentence and adjust the error results to the right results: huixing and No.85 We found many cases like this one where important named entities are correctly recognized with the help of our context-aware rescoring approach. This is beneficial to downstream tasks.

        \begin{table}[htbp]
        \caption{Examples of context-aware lattice rescoring results}
        \label{tab:egs}
        \begin{tabular}{llll}
        \toprule[1pt]
        \multicolumn{4}{c}{\emph{Reference}} \\ \hline
        \multicolumn{2}{l}{\begin{tabular}[c]{@{}l@{}}A:Where do you want to meet, huixin garden? \\ A:Hello, may No.85 help you\end{tabular}} &
          \multicolumn{2}{l}{\begin{tabular}[c]{@{}l@{}}B:\textbf{huixin} garden \\ B:\textbf{No.85},Yes\end{tabular}} \\ \hline
        \multicolumn{4}{c}{\emph{RNNLM rescoring}} \\ \hline
        \multicolumn{2}{l}{\begin{tabular}[c]{@{}l@{}}A:Where do you want to meet, huixin garden?\\ A:Hello, may No.85 help you\end{tabular}} &
          \multicolumn{2}{l}{\begin{tabular}[c]{@{}l@{}}B:\textbf{\textcolor[rgb]{1,0,0}{huixing}} garden\\B:\textbf{\textcolor[rgb]{1,0,0}{No.89}},Yes\end{tabular}} \\ \hline
        \multicolumn{4}{c}{\emph{Context-aware rescoring}} \\ \hline
        \multicolumn{2}{l}{\begin{tabular}[c]{@{}l@{}}A:Where do you want to meet, huixin garden?\\ A:Hello, may No.85 help you\end{tabular}} &
          \multicolumn{2}{l}{\begin{tabular}[c]{@{}l@{}}B:\textbf{huixin} garden\\ B:\textbf{No.85},Yes\end{tabular}} \\ \bottomrule[1pt]
        \end{tabular}
        \end{table}

\section{Conclusions}
In this work, we propose a context-aware lattice rescoring methods for RNNLMs to capture topic effects and long-distance triggers for conversational speech recognition. CER improvements have been achieved on both Yispeech and HKUST corpora. We have achieved up to 6\% improvement on Yispeech. For all four test sets, the accuracy of essential words recognition has been improved. We show that it is sufficient to use the context-aware method on RNNLM and lattice rescoring and use the similarity score to rescore lattices selectively. In the future, we will explore the effect of a more extended range of contextual connections, as well as study effective methods of incorporating additional information such as intent information to improve the effectiveness of conversational speech recognition.

\bibliographystyle{IEEEtran}

\bibliography{mybib}


\end{document}